\documentclass[draft,grl]{agutex}

\usepackage{lineno}

\usepackage[dvips]{graphicx}
\usepackage{ulem}

\setkeys{Gin}{draft=false}
\authorrunninghead{Suweis et al.}

\titlerunninghead{{\sc Structure and Controls of the GVWTN}}

\begin{document}

%
%

\title{Structure and Controls of the Global Virtual Water Trade Network}


%
%

\authors{S. Suweis\altaffilmark{1,2}, M. Konar\altaffilmark{2}, C. Dalin\altaffilmark{2}, N. Hanasaki\altaffilmark{3}, A. Rinaldo\altaffilmark{1,4},\\ and I. Rodriguez-Iturbe\altaffilmark{2}}

\altaffiltext{1}{Laboratory of Ecohydrology ECHO/IIE/ENAC,
\'Ecole Polytechnique F\'ed\'erale, Lausanne, 1015  Switzerland.}
\altaffiltext{2}{Department of Civil and Environmental Engineering, Princeton University,
Princeton, NJ 08544, USA}
\altaffiltext{3}{National Institute for Environmental Studies, 16-2 Onogawa, Tsukuba, Ibaraki 305-8506 Japan}

\altaffiltext{4}{Dipartimento IMAGE, University of Padua, Padova 35131, Italy}

%
%
%

\begin{abstract}
Recurrent or ephemeral water shortages are a crucial
global challenge, in particular because of their impacts
on food production. The global character of this challenge is
reflected in the trade among nations of virtual water, i.e. the
amount of water used to produce a given commodity. We build,
analyze and model the network describing the transfer of virtual
water between world nations for staple food products. We find that
all the key features of the network are well described by a model
that reproduces both the topological and weighted properties of
the global virtual water trade network, by assuming as sole
controls each country's gross domestic product and yearly rainfall
on agricultural areas. We capture and quantitatively
describe the high degree of globalization of water trade and show
that a small group of nations play a key role in the connectivity
of the network and in the global redistribution of virtual water.
Finally, we illustrate examples of prediction of the structure of the network under future
political, economic and climatic scenarios, suggesting that the
crucial importance of the countries that trade large volumes of water will be
strengthened.

\end{abstract}

%
%

\begin{article}

%
%

\section{Introduction}\label{I}

Food production is by far the most freshwater-consuming process
(80\% of the total world water resources \citep{rost2008}). Due to
population growth and economic development, water shortage is thus
subject to increasing pressure at local and global scales. Several
studies have recently focused on the issues of globalization of
water \citep[e.g.][]{hoekstra2002,chapagain2006,Dodorico2010},
using the concept of virtual water (VW) \citep{allan1993}. They
have highlighted the importance of tackling water management
problems not only at the basin or country scales, but
rather through a worldwide perspective
\citep{hoekstra2008}. Nevertheless a  complete statistical
characterization describing the network of VW transfers on a
global scale has only started \citep{konar2010} and a model to
explain such characteristics is still lacking.

This Letter deals with a novel theoretical network model
that robustly describes topological and weighted properties of the
global VW trade network (GVWTN) for the characterization of the VW
flows of $5$ major crops (barley, corn, rice, soy, and wheat)
and $3$ livestock products (beef, poultry, and pork). The VW content of such
commodities are calculated for each nation using a
state-of-the-art global water resources model
\citep{hanasaki2008b,hanasaki2010}, at a spatial scale of
$0.5^{o}\times 0.5^{o}$. Combining the model outputs with the data
of the international trade of food products referred to the year
2000 \citep{FAO}, the VW flows among nations are obtained (see
\cite{konar2010} and Auxiliary Materials for details) and compared
with model results. Of particular interest is deemed our
predictive use of the model to analyze the impact of future
scenarios of social, economic, and climate change.

\section{Structure of the GVWTN}\label{S}
Here we briefly present and interpret key statistical
characterizations of the GVWTN (addressed in \cite{konar2010}).
For simplicity we discuss here the undirected network case, where
$W$ is a symmetric matrix whose elements ($w_{ij}=w_{ji}$)
represent the total volume of VW exchanged between countries
(nodes) and obtained by summing the corresponding import and
export fluxes (Figure 1). For mathematical details and
analysis of the directed network see Auxiliary Materials.

The global topology of a network is described by its
degree probability density function (pdf) $p(k)$, i.e. $p(k) dk$
is the probability that the degree of a given node is $k$
\citep{newmanbarabasiwatts} (Figure 2a).
It provides the number of edges connected to a given node
regardless of the identity of the neighbors. To investigate how
nodes are connected, we study the average nearest neighbor degree
$k_{nn}$ which shows a tendency of the nodes with high degree to
provide connectivity to small degree nodes (Figure 2c). Such trend
(known as disassortative behavior) denotes, differently from
purely random networks, non-trivial nodal degree correlations
\citep{newman2002}. Another interesting indicator is the local
clustering coefficient $C_i$ ($0 \leq C_i \leq 1$) which describes
the ability of node $i$ to form cliques, i.e. triangles of
connected nodes. Figure 2d shows that poorly connected nations $i$
tends to form connected trading food sub-markets ($C_i \approx
1$). On the contrary, high degree vertices $j$ connect otherwise
disconnected regions ($C_j \ll 1$). The average clustering
coefficient is very high ($\bar{C}=0.747$) and the graph has,
in analogy to many real networks \citep{newmanbarabasiwatts}, an
average node-to-node topological distance ($d_{nn}$) smaller than
5 ($d_{nn}=4$) \citep{konar2010}.
The GVWTN thus exhibits a small-world network behavior
\citep{watts_and_strogatz}, providing a quantitative measure of
the globalization of water resources \citep{hoekstra2008,Dodorico2010}.

The hydrological features of the network are given by its weighted
properties. The total volume imported and exported by
nation $i$ is quantified by its strength $s_i$, defined as the
total VW volume exchanged by node $i$. The strength distribution
shows a heavy-tailed pdf suggesting high heterogeneity of the
volumes of traded VW (Figure 2b): only 4\% of the total number of
links accounts for 80\% of the total flow volume
\citep{konar2010}, indicating established bonds among countries
that rule the main fluxes in the GVWTN (Figure 1). Strengths
between neighboring nodes are correlated. In fact, the average
nearest neighbor strength $s_{nn}$ \citep{serrano2008} displays a
decreasing trend as a function of $s$ (Figure 2e).
Strength-strength correlations disentangled from degrees
($s^W_{nn}$) \citep{serrano2008} are not significant, i.e.
$s^W_{nn}$ does not depend on $s$. A power-law relation $s\sim
k^b$ with exponent $b=2.60$ \citep{konar2010} indicates a
non-trivial correlation between node degrees and strengths
\citep{barrat2004}. The above suggests that we live in a global
water world where, on average, the export of VW from few water
rich countries increases the food locally available to the
connected nations. At the same time, there exist preferential VW
routes, mainly driven by geographical, political and economical
factors, through which most of the VW volume flows.

\section{Controls of the GVWTN}\label{C}
The complexity of all factors (political, economical and
environmental) involved in shaping the GVWTN structure is
remarkable, and calls for investigating whether key variables and
linkages exist through which the emerging structural properties of
the network could be revealed. We have developed a model that
allows us to describe concisely all the above features of the
GVWTN. Specifically, we assume that the topological and weighted
features of the network can be determined, respectively, by two
external characteristics of each node: namely, the gross domestic
product \citep{WB} ($GDP$) and the (average) yearly rainfall
[mm/yr] on agricultural area [km$^2$] (denoted by $RAA$
[mm$\cdot$km$^2$/yr]).

Toward this end, each of the 184 nodes is assigned a normalized
value of the $GDP$ ($x$) and $RAA$ ($y$) based on data from 2000
\citep{UN,WB} (i.e., $x_i=GDP_i/\sum_{j=1}^NGDP_j$,
$y_i=RAA_i/\sum_{j=1}^NRAA_j)$. We refer to these variables as
fitness (or hidden) variables
\citep[e.g.][]{bianconi2001,caldarelli2002,boguna2003,garlaschelli2004,Park2004}.
They measure the relative importance of the vertices in the GVWTN.
$GDP$ and $RAA$ are assumed to be good candidates to
explain the structure of the GVWTN. In fact the country $GDP$ is
closely related to its trade activity \citep{garlaschelli2004},
while volumes of VW traded depend on the amount of crops and meat
produced in that country, that in turn depends on the $RAA$. A
good agreement between data and model results proves these facts.
The fitness network-building algorithm consists of the
following steps: $a)$ we connect every couple of vertices, $i$,
$j$, (with $i\neq j$) with a probability $p(x_i,x_j)=\sigma x_i
x_j /(1+\sigma x_i x_j)$; $b)$ we assign to each link between $i$
and $j$ a weight $\langle w_{ij} \rangle$ with value given by
$q(y_i,y_j)=\eta y_i y_j$. The parameters of the model are
$\sigma$ and $\eta$ and they are determined by the compatibility
conditions: $\frac{1}{2}\sum_{i}\sum_{j\neq i}p(x_i; x_j)=L$ and
$\frac{1}{2}\sum_{i}\sum_{j\neq i}q(y_i,y_j)=\Phi$, where $L$ is
the total number of edges in the network and $\Phi$ the total
flux. No tuning of the parameters is carried out and all results
presented here correspond to the above choice for $\sigma$ and
$\eta$. For details on the model and its generalization to the
directed network case see Appendix and Auxiliary Materials.

The model predicts that for each country the number of
food trade partners grows non linearly with its $GDP$ (similarly
to the world trade web \citep{garlaschelli2004}), while the total
exported and imported VW is found proportional to the $RAA$. Exact
results are obtained on the properties related to the node
strengths (see Appendix for details). The analytical
results match closely the empirical ones. In particular we find
that $\langle s(y) \rangle =N \eta \langle y \rangle y$ (where $
\langle\cdot\rangle$ represents the ensemble average and $s(y)$ is
the strength of a node associated with the value $y$ of the
normalized $RAA$), $\langle s^W_{nn}\rangle=N \eta\gamma^2
\Gamma[1+2/\beta]$ (where $\Gamma[\cdot]$ is the complete Gamma
function) and that the pdf of the node strength is
$p(s)=\mathcal{C}s^{\beta-1}e^{(-\frac{s}{\gamma\eta})^\beta}$,
where $\mathcal{C}=\beta/(\gamma\eta)^{\beta}$ with
$\beta\approx0.482$ and $\gamma\approx0.002$ parameters related to
the pdf of $y$ (see Auxiliary Materials for details). From the
empirical relation $s=a k^b$ it can be shown that the cumulative
degree pdf decays exponentially as
$P_{>}(k)=\int_k^{\infty}p(k')dk'= e^{-\alpha k^{\beta b}}$, with
$\alpha=(\gamma\sigma/a)^{-\beta}$ and $\beta\cdot b\approx1.25$.
Figures 2 and 3 summarize all the results discussed above. They
show the excellent agreement of the fitness model with the
empirical data.

\section{Future Scenarios}\label{F}
Our theoretical framework is suitable to investigate future
scenarios of the GVWTN structure. To this aim, we evaluate
estimates of the annual rainfall for 2030-2050 from the A2 socio-economic
scenario of the World Climate Research Programmes (WCRPs) Coupled Model Intercomparison
Project Phase 3 (CMIP3) multi-model dataset \citep{CCS}. The
spatial mean is then calculated for each country in the network
over this time horizon. Then by using published projections of the
$GDP$ and agricultural area \citep{FAO2,Rosegrant2009} for 2030,
we build the fitness functions $p(x^T_i,x^T_j)=\sigma'
x^T_ix^T_j/(1+\sigma' x^T_ix^T_j)$ and $q(y^T_i,y^T_j)=\eta' y^T_i
y^T_j$, where $x^T$ and $y^T$ are the projections of the fitness
variables at year $T=2030$. The parameters $\sigma'$ and $\eta'$
are to be determined by the future total number of connections
$L'$ and flux $\Phi'$. In our simulation we assumed that $L'=L$
and $\Phi'=\Phi$, but in general they may be part of the scenarios
under study. All A2 climate change scenarios
\citep{CCS} yield a decrease in rainfall at a global scale, but
the total arable land is predicted by \citep{FAO2} to increase
around 1\%, thereby leading to an increase of the total $RAA$.
Figure 4 summarizes the results of the structure of the GVWTN
under the driest climate change scenario. We find that the
structure of the GVWTN topology is robust with respect to these particular
scenarios. A heavier tail in the strengths pdf is observed
suggesting a rich-gets-richer phenomenon
\citep{newmanbarabasiwatts}, where the nodes with large strengths
benefit from the changes in $RAA$, becoming even stronger. We also
find that the exponent in  the node independent relation $s^*\sim {k^*}^q$,
(where the vectors $k^*$ and $s^*$ are the sorted degrees and strengths in the GVWTN)
increases from $q=2.69\pm0.03$ ($R^2=0.982$) to $q=2.77\pm0.02$ ($R^2=0.986$) (see Figure 3c and inset Figure 4b).
These results suggest that economic and climatic future scenarios will likely enhance the
globalization of water resources, giving to water-rich countries
even more inroad for reaching poorly connected nodes. At the same
time, the observed rich-gets-richer phenomenon will intensify the
reliance of most of the nations on the few VW hubs. As a consequence it will
reduce the ability of the GVWTN to respond to disturbances whose
impact may be dramatic when the VW trade supports carrying
capacities beyond those supported by local resources
\citep{Dodorico2010}. Finally our study highlights how
agricultural land management may indeed remarkably impact the
future structure of the GVWTN.

Our work opens new quantitative and predictive perspectives in the
study of stability and complexity of the GVWTN coupled to social,
economic and political processes related to the international food
trade. Ongoing research incorporates scenarios where $L'$ and $\Phi'$ are
different from values of the year 2000 and reflect the evolving and dynamic character of the global  network.


%
%

\section*{Appendix}

Our modelling scheme for the properties of the GVWTN employs: i) a
function describing the topological properties of the network, and
ii) a function, independent of i), characterizing its weights. The
functional shape of $p(x_i,x_j)$, the probability that node $i$
and $j$ -- endowed respectively with fitness $x_i$ and $x_j$ --
are connected, is found through an entropy optimization principle
and by imposing that all graphs with degree sequence
$\{k_l\}_{l=1,2,..}$ appear in our ensemble with equal probability
(see Auxiliary Materials and \cite{Park2004} for details). The
result is
\begin{equation}\label{pij}
    p(x_i,x_j)=\frac{\sigma x_i x_j}{1+\sigma x_i x_j}.
\end{equation}
The key assumption is that fitness variable $x_i$ is assumed to be
the external quantity $x_i=GDP_i/(\sum_j GDP_j)$, determining the
topological importance of node $i$ by driving the number of its
connections. From Eq. (\ref{pij}) one computes all topological
properties (defined in the Auxiliary Materials): the node degree $
\langle k_i \rangle =\sum_{j\neq i}^N p(x_i,x_j)$; the average
degree of the nearest neighbors
\begin{equation}\label{knnF}
   \langle k_{nn,i} \rangle =\frac{\sum_{j\neq i}^N\sum_{l\neq j} p(x_i,x_j)p(x_j,x_l)}{ \langle k_i \rangle},
\end{equation}
and the local clustering coefficient:
\begin{equation}\label{CF}
   \langle C_{i} \rangle =\frac{\sum_{j\neq i}^N\sum_{l\neq j,i}^N\ p(x_i,x_j)p(x_j,x_l)p(x_l,x_i)}{(\langle k_i \rangle-1)\langle k_i \rangle}.
\end{equation}
A function $q(y_i,y_j)$ assigns the average weight $\langle
w_{ij}\rangle$ to the link connecting $i$ to $j$ as a function of
the fitness variables $y$. We interpret $\langle w_{ij}\rangle$ as
a rank associated with the assigned link between two nodes $i$ and
$j$ and its importance. By generalizing the concept of weighted
configuration model \citep{serrano2005,garlaschelli2009}, our null
hypothesis for $q(y_i,y_j)$ is:
\begin{equation}\label{qij}
    q(y_i,y_j)=\eta y_i y_j,
\end{equation}
where $\eta$ is the parameter controlling the total flux of the
network. We choose as fitness variable $y$ the normalized rainfall
on agricultural area $y_i=RAA_i/\sum RAA_i$.

Given the simple functional shape of Eq. (\ref{qij}), if an
analytical approximation for the distribution of $y$ exists, exact
results can be obtained on the properties involving node
strengths. We find that the empirical cumulative distribution of
$y$ is well fitted ($R^2 \approx 0.998$) by a stretched
exponential $\rho_{>}(y)=\exp \left(-(\frac{y}{\gamma})^\beta
\right)$. Then using the continuum approximation
\citep{caldarelli2002}, we obtain $\langle s(y) \rangle=N
\int_0^{\infty} q(y,z)\rho(z)dz=N\langle y\rangle\eta y=N F(y)$,
and for large enough $N$ one has: $p(s)=\rho[N
F^{-1}(s/N)]\frac{d}{ds}F^{-1}(s/N)=\frac{1}{\eta}\rho(s/\eta)$,
yielding:
\begin{equation}\label{Ps}
    P_>(s)=e^{-(\frac{s}{\eta \gamma})^\beta}.
\end{equation}
Finally, the strength-strength correlation ia obtained as:
\begin{equation}\label{SWnn}
    \langle s^W_{nn}(y)\rangle=\frac{N \int_0^{\infty}q(y,z)\langle s(z) \rangle\rho(z)dz}{\langle s(y) \rangle} =N \eta\gamma^2 \Gamma[1+2/\beta],
\end{equation}
and it is found that it that does not depend on $y$. Although we
are not able to repeat the same procedure for the fitness variable
$x$, a qualitative analytical behavior for the distribution of the
node degree can be obtained by using the empirical relation $s=a
k^b$ through a derived distribution approach, i.e.
$p(k)dk=p(s)ds$. We then find:
\begin{equation}\label{Pk}
    P_>(k)=e^{-(\frac{a}{\eta \gamma})^\beta k^{b\beta}},
\end{equation}
which is a compressed exponential distribution, confirming the
exponential-like tail observed from the empirical analysis of the
degree distribution. Further details are in the Auxiliary
Materials.


%

%
\begin{acknowledgments}
IRI, MK and CD gratefully acknowledge the support of the James S. McDonnell Foundation (Grant 220020138). We acknowledge the Program for Climate Model Diagnosis and Intercomparison (PCMDI) and the WCRP Working Group on Coupled Modelling (WGCM) for making available the WCRP CMIP3 multi-model dataset supported by the Office of Science, U.S. Department of Energy. SS and AR gratefully acknowledge the support provided by the ERC Advanced Grant RINEC-227612 and by the SFN/FNS project $200021 124930/1$.
\end{acknowledgments}


\clearpage
\newpage

\end{article}
\begin{figure}[h!]
\begin{center}
\includegraphics[width=8cm]{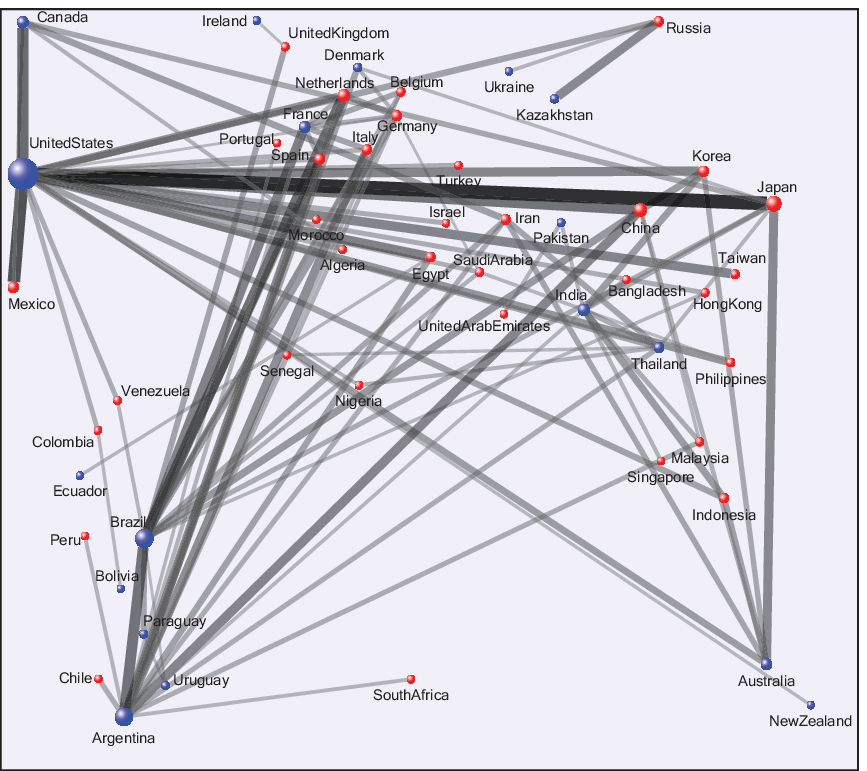}
\end{center}
\caption{Backbone of the global Virtual Water Trade Network
(GVWTN). Only 4\% of the total number of links accounting for 80\%
of the total flow volume are shown. Resulting isolated nodes are
consequently removed. The blue nodes represent the net exporter
nations, while the red ones are the net importers. The weights of
the links are color-coded by the grayscaling in the edge's colors
(black is the link carrying the highest volume of VW.)}
\end{figure}

\clearpage
\newpage

\begin{figure}[h!]
\begin{center}
\includegraphics[width=8cm]{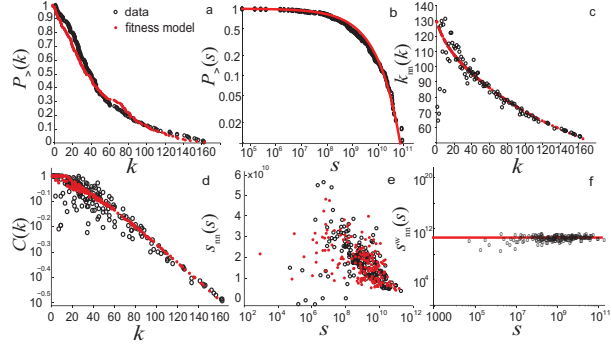}
\end{center}
\caption{Topological and weighted properties of the GVWTN compared
with the results of the fitness model (red line). a-b)
Cumulative pdf of the node's degree $P_{>}(k)$ in linear
scale and node strength $P_{>}(s)$ in log-log scale; c-d) average
nearest neighbors degree $k_{nn}$ and cluster coefficient $C$ as a
function of the nodes degree $k$; e-f) average nearest neighbors
strength $s_{nn}$ and strengths-strengths correlation
$s^{W}_{nn}(s)\sim const$ in semilog-$x$ and log-log scale, respectively.}
\end{figure}

\clearpage
\newpage

\begin{figure}
\begin{center}
\includegraphics[width=8cm]{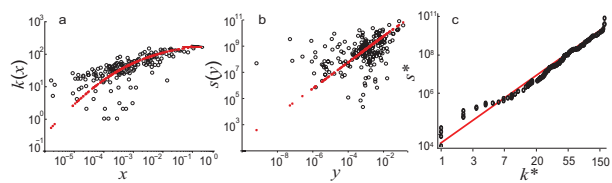}
\end{center}
\caption{Comparison between empirical (black dots) and model
results (red dots): a) The relationship between nodal degrees  $k$
and normalized $GDP$ $x$ shows how on average the number
of connections is an increasing function of the nation's $GDP$;
b) The relationship between node strength $s$ and normalized
$RAA$, $y$ i.e. $s(y) = N \eta\langle y\rangle y$; c) The
\textbf{node independent} relationship between strengths $s^*$ and degrees $k^*$
in the GVWTN ($s^* \sim {k^*}^q$). The red line represents the best fit obtained
from $k$ and $s$
generated by the fitness model. In this case, we find $ q = 2.69
\pm 0.03$ ($R^2=0.98$). All the plots are in log-log scale.}
\end{figure}

\clearpage
\newpage

\begin{figure}[h!]
\begin{center}
\includegraphics[width=8cm]{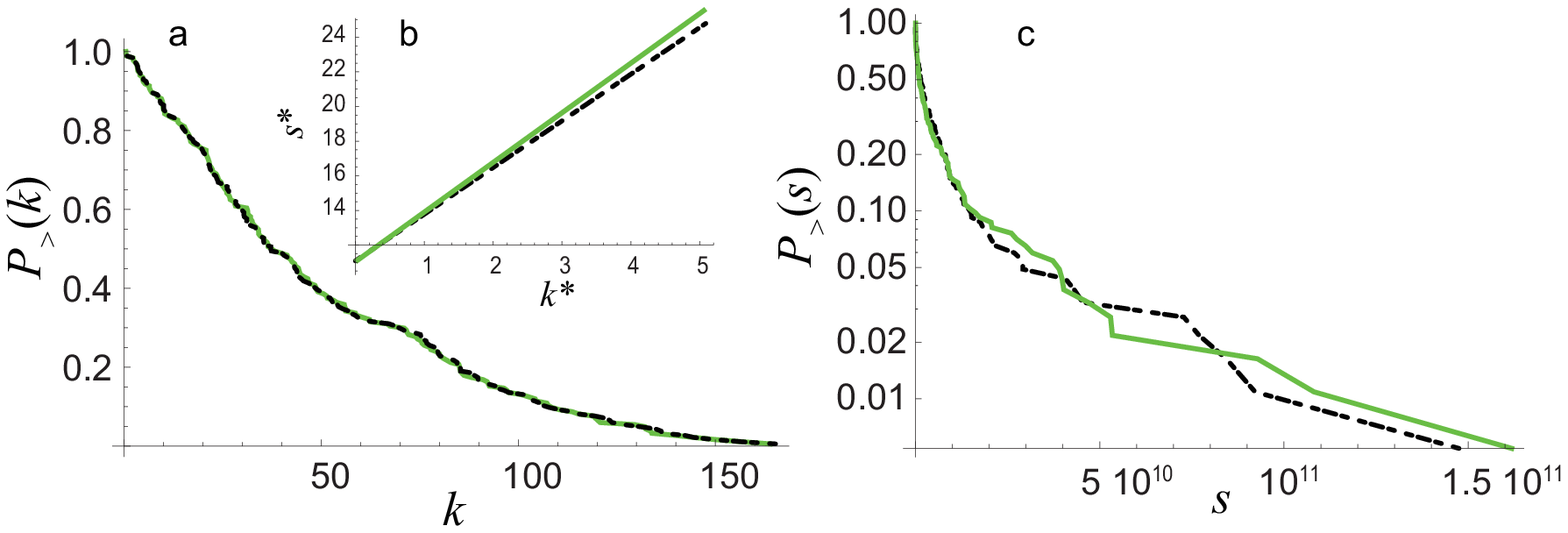}
\end{center}
\caption{An example of predictive application of the fitness model
of GVWTN for the driest case scenario:  comparison between the
global properties of the network for the year 2000 (dashed black)
with those predicted by the fitness model for the year 2030 (green
line): a) Cumulative degree pdf. Inset b): Relationships between
sorted strengths and degrees); c) Cumulative pdf of the nodal
strengths.}
\end{figure}


\begin{thebibliography}{27}
\bibliographystyle{agu}
\providecommand{\natexlab}[1]{#1}
\expandafter\ifx\csname urlstyle\endcsname\relax
  \providecommand{\doi}[1]{doi:\discretionary{}{}{}#1}\else
  \providecommand{\doi}{doi:\discretionary{}{}{}\begingroup
  \urlstyle{rm}\Url}\fi

\bibitem[{\textit{Allan}(1993)}]{allan1993}
Allan, T. (1993), Fortunately there are substitutes for water: otherwise our
  hydropolitical futures would be impossible, \textit{Proceedings of the
  Conference on Priorities for Water Resources Allocation and Management},
  \textit{2}, 13--26.

\bibitem[{\textit{Barrat et~al.}(2004)\textit{Barrat, Barthelemy,
  Pastor-Satorras, and Vespignani}}]{barrat2004}
Barrat, A., M.~Barthelemy, R.~Pastor-Satorras, and A.~Vespignani (2004), The
  architecture of complex weighted networks, \textit{P. Natl. Acad. Sci. Usa},
  \textit{101}(11), 3747--3752.

\bibitem[{\textit{Bianconi and Barabasi}(2001)}]{bianconi2001}
Bianconi, G., and A.~Barabasi (2001), Bose-einstein condensation in complex
  networks, \textit{Phys. Rev. Lett.}, \textit{89}(25), 5663.

\bibitem[{\textit{Boguna and Pastor-Satorras}(2003)}]{boguna2003}
Boguna, M., and R.~Pastor-Satorras (2003), Class of correlated random networks
  with hidden variables, \textit{Phys. Rev. E}, \textit{68}, 036,112.

\bibitem[{\textit{Caldarelli et~al.}(2002)\textit{Caldarelli, Capocci,
  De~Los~Rios, and Munoz}}]{caldarelli2002}
Caldarelli, G., A.~Capocci, P.~De~Los~Rios, and M.~A. Munoz (2002), {Scale-Free
  Networks from Varying Vertex Intrinsic Fitness}, \textit{Phys. Rev. Lett.},
  \textit{89}(25), 258,702.

\bibitem[{\textit{Chapagain et~al.}(2006)\textit{Chapagain, Hoekstra, and
  Savenije}}]{chapagain2006}
Chapagain, A.~K., A.~Hoekstra, and H.~Savenije (2006), Water saving through
  international trade of agricultural products, \textit{Hydrol. Earth Syst.
  Sci.}, \textit{10}, 455--468.

\bibitem[{\textit{D'Odorico et~al.}(2010)\textit{D'Odorico, Laio, and
  Ridolfi}}]{Dodorico2010}
D'Odorico, P., F.~Laio, and L.~Ridolfi (2010), Does globalization of water
  reduce societal resilience to drought?, \textit{Geophys. Res. Lett.},
  \textit{37}, L13,403.

\bibitem[{\textit{{FAO}}(2000{\natexlab{a}})}]{FAO}
{FAO} (2000{\natexlab{a}}), Food trade data, \textit{{www.faostat.fao.org}}.

\bibitem[{\textit{{FAO}}(2000{\natexlab{b}})}]{FAO2}
{FAO} (2000{\natexlab{b}}), World agriculture: towards 2030/2050,
  \textit{http://ftp.fao.org/docrep/fao/009/a0607e/a0607e00.pdf}.

\bibitem[{\textit{Fonseca et~al.}({2009})\textit{Fonseca, Narrod, Rosegrant,
  Fernandez, Sinha, and Alder}}]{Rosegrant2009}
Fonseca, J., C.~Narrod, M.~W. Rosegrant, M.~Fernandez, A.~Sinha, and J.~Alder
  ({2009}), {Looking into the future for agriculture and AKST},
  \textit{{IAASTD}}, \textit{{5}}, {307--37}.

\bibitem[{\textit{Garlaschelli and Loffredo}(2004)}]{garlaschelli2004}
Garlaschelli, D., and M.~I. Loffredo (2004), Fitness-dependent topological
  properties of the world trade web, \textit{Phys. Rev. Lett.},
  \textit{93}(18), 188,701.

\bibitem[{\textit{Garlaschelli and Loffredo}(2009)}]{garlaschelli2009}
Garlaschelli, D., and M.~I. Loffredo (2009), Generalized bose-fermi statistics
  and structural correlations inweighted networks, \textit{Phys. Rev. Lett.},
  \textit{102}, 038,701.

\bibitem[{\textit{Hanasaki et~al.}(2008)\textit{Hanasaki, Kanae, Oki, Masuda,
  Motoya, Shirakawa, Shen, and Tanaka}}]{hanasaki2008b}
Hanasaki, N., S.~Kanae, T.~Oki, K.~Masuda, K.~Motoya, N.~Shirakawa, Y.~Shen,
  and K.~Tanaka (2008), An integrated model for the assessment of global water
  resources - part 2: Applications and assessments, \textit{Hydrol. Earth Syst.
  Sci.}, \textit{12}(3-4), 1027--1037.

\bibitem[{\textit{Hanasaki et~al.}(2010)\textit{Hanasaki, Inuzuka, Kanae, and
  Oki}}]{hanasaki2010}
Hanasaki, N., T.~Inuzuka, S.~Kanae, and T.~Oki (2010), An estimation of global
  virtual water flow and sources of water withdrawal for major crops and
  livestock products using a global hydrological model, \textit{J. Hydrol.},
  \textit{384}(3-4), 232--244.

\bibitem[{\textit{Hoekstra and Chapagain}(2008)}]{hoekstra2008}
Hoekstra, A., and A.~K. Chapagain (2008), \textit{Globalization of Water},
  Blackwell.

\bibitem[{\textit{Hoekstra}(2002)}]{hoekstra2002}
Hoekstra, A.~Y. (2002), Virtual water trade: Proceedings of the international
  expert meeting on virtual water trade, in \textit{Value of Water Researh
  Report}, 12, UNESCO-IHE, Delft.

\bibitem[{\textit{Konar et~al.}(2011)\textit{Konar, Dalin, Suweis, Hanasaki,
  Rinaldo, and Rodriguez-Iturbe}}]{konar2010}
Konar, M., C.~Dalin, S.~Suweis, N.~Hanasaki, A.~Rinaldo, and
  I.~Rodriguez-Iturbe (2011), Water for food: The global virtual water trade
  network, \textit{Accepted in Water Resour. Res.}

\bibitem[{\textit{Meehl et~al.}(2007)\textit{Meehl, Covey, Delworth, Latif,
  McAvaney, Mitchell, Stouffer, and Taylor}}]{CCS}
Meehl, G.~A., C.~Covey, T.~Delworth, M.~Latif, B.~McAvaney, J.~F.~B. Mitchell,
  R.~J. Stouffer, and K.~E. Taylor (2007), {The WCRP CMIP3 MULTIMODEL DATASET:
  A New Era in Climate Change Research}, \textit{Bull. Amer. Met. Soc.},
  \textit{88}, 1383--1394.

\bibitem[{\textit{Newman}({2002})}]{newman2002}
Newman, M. ({2002}), {Assortative Mixing in Networks}, \textit{{Phys. Rev.
  Lett.}}, \textit{{89}}({20}), {208,701}.

\bibitem[{\textit{Newman et~al.}(2006)\textit{Newman, Barabasi, and
  Watts}}]{newmanbarabasiwatts}
Newman, M., A.~L. Barabasi, and D.~J. Watts (2006), \textit{The Structure and
  Dynamics of Networks}, Princeton University Press.

\bibitem[{\textit{Park and Newman}(2004)}]{Park2004}
Park, J., and M.~E.~J. Newman (2004), Statistical mechanics of networks,
  \textit{Phys. Rev. E}, \textit{70}, 066,117.

\bibitem[{\textit{Rost et~al.}(2008)\textit{Rost, Gerten, Bondeau, Lucht,
  Rohwer, and Schaphoff}}]{rost2008}
Rost, S., D.~Gerten, A.~Bondeau, W.~Lucht, J.~Rohwer, and S.~Schaphoff (2008),
  Agicultural green and blue water consumption and its influence on the global
  water system, \textit{Water Resour. Res.}, \textit{44}, W09,405,
  doi:10.1029/2007WR006,331.

\bibitem[{\textit{Serrano}(2008)}]{serrano2008}
Serrano, M.~A. (2008), Rich-club vs rich-multipolarization phenomena in
  weighted networks, \textit{Phys. Rev. E}, \textit{78}, 026,101.

\bibitem[{\textit{Serrano and Boguna}(2005)}]{serrano2005}
Serrano, M.~A., and M.~Boguna (2005), {Weighted Configuration Model},
  \textit{{AIP Conf. Proc.}}, \textit{776}, 101.

\bibitem[{\textit{{United Nations}}(2010)}]{UN}
{United Nations} (2010), United nation (statistic division),
  \textit{http://unstats.un.org/unsd/environment/waterresources.htm}.

\bibitem[{\textit{Watts and Strogatz}(1998)}]{watts_and_strogatz}
Watts, D.~J., and S.~H. Strogatz (1998), Collective dynamics of small-world
  networks, \textit{Nature}, \textit{393}(6684), 440--442.

\bibitem[{\textit{{World Bank}}(2010)}]{WB}
{World Bank} (2010), World bank data,
  \textit{http://data.worldbank.org/indicator}.

\end{thebibliography}
\end{document}